\pdfoutput=1

\documentclass[11pt]{article}

\usepackage[final]{ACL2023}

\usepackage{times}
\usepackage{latexsym}
\usepackage{graphicx}
\usepackage{hyperref}
\usepackage{fancyvrb} 
\usepackage{tcolorbox} 
\usepackage{booktabs}
\usepackage{svg}
\usepackage[T1]{fontenc}

\usepackage[utf8]{inputenc}

\usepackage{microtype}

\usepackage{inconsolata}

\title{Nyay-Darpan: Enhancing  Decision Making Through Summarization and Case Retrieval for Consumer Law in India}

\author{
Swapnil Bhattacharyya\textsuperscript{1} \quad
Harshvivek Kashid\textsuperscript{1} \quad
Shrey Ganatra\textsuperscript{1} \quad
Spandan Anaokar\textsuperscript{1} \\
Shruti Nair\textsuperscript{2} \quad
Reshma Sekhar\textsuperscript{2} \quad
Siddharth Manohar\textsuperscript{2} \quad
Rahul Hemrajani\textsuperscript{2} \\
\textbf{Pushpak Bhattacharyya}\textsuperscript{1} \\
\textsuperscript{1}Indian Institute of Technology Bombay \quad
\textsuperscript{2}National Law School of India University, Bangalore \\
\texttt{\{ganatrashrey2002, harshvivek14, spandananao, pushpakbh\}@gmail.com} \\
\texttt{swapnilbhyya@cse.iitb.ac.in}
}

\begin{document}
\maketitle
\begin{abstract}
AI-based judicial assistance and case prediction have been extensively studied in criminal and civil domains, but remain largely unexplored in consumer law, especially in India. In this paper, we present Nyay-Darpan, a novel two-in-one framework that (i) summarizes consumer case files and (ii) retrieves similar case judgements to aid decision-making in consumer dispute resolution. Our methodology not only addresses the gap in consumer law AI tools but also introduces an innovative approach to evaluate the quality of the summary. The term 'Nyay-Darpan' translates into 'Mirror of Justice', symbolizing the ability of our tool to reflect the core of consumer disputes through precise summarization and intelligent case retrieval. Our system achieves over 75 percent accuracy in similar case prediction and approximately 70 percent accuracy across material summary evaluation metrics, demonstrating its practical effectiveness. We will publicly release the Nyay-Darpan framework and dataset to promote reproducibility and facilitate further research in this underexplored yet impactful domain.
\end{abstract}

\section{Introduction}
The increasing complexity of consumer law and the rapid expansion of legal case data have introduced several challenges in consumer law forums. Legal professionals face not only the manual effort required to analyze extensive case files but also additional obstacles, including the ambiguity in sector classification, inconsistent document structures, and the lack of domain-specific datasets that can support consumer law summarization and retrieval efficiently. Furthermore, the risk of hallucinations in LLM outputs and the jurisdictional variability of legal reasoning add to the difficulty of automating reliable legal decision support.

To address these challenges, Artificial Intelligence (AI) and Natural Language Processing (NLP) techniques have gained significant attention for automating legal text analysis \cite{katz2023naturallanguageprocessinglegal}. Prior works have employed machine learning models for legal case summarization, enhancing legal research efficiency and accessibility, with both abstractive and extractive summarization explored in the Indian legal context \cite{shukla-etal-2022-legal}.

The application of LLMs in the legal domain has primarily focused on legal judgment prediction \cite{shui-etal-2023-comprehensive}, case summarization, retrieval of prior cases, and identification of legal statutes \cite{joshi-etal-2024-il, feng-etal-2024-legal}. Although legal LLMs have been developed \cite{zhou2024lawgptchineselegalknowledgeenhanced}, none specifically target consumer law in India. Predicting similar cases remains a crucial task for legal practitioners and judges to cite appropriate precedents, as highlighted by \citet{10255647} in the context of civil, criminal, and human rights domains. Decision assist tools can substantially alleviate cognitive burden by providing succinct and contextually relevant summaries, enabling even non-experts to comprehend complex legal outcomes more easily \cite{jiang-etal-2024-leveraging}. Relevant advances in this field include U-creat for unsupervised case retrieval \cite{joshi-etal-2023-u} and graph-based retrieval techniques \cite{kipf2017semisupervisedclassificationgraphconvolutional}.

In September 2023, more than 5.45 lakh cases remained pending in consumer commissions\footnote{\url{https://www.pib.gov.in/}} of India, emphasizing the need for decision-assist tools that can accelerate legal processes. In this work, we propose an AI-powered decision-assist tool for consumer law forums and consumers that integrates material summarization and similar case prediction, employing sector-based classification and a combination of lexical and semantic similarity to retrieve relevant precedents efficiently. The tool is also potentially useful for law firms and lawyers practicing in consumer cases. \\
Our contributions are as follows:
\begin{enumerate}
\item \textbf{Consumer Decision Assist Tool (Summarizer)}, A part-wise, CoT-prompted summarizer tailored for Indian consumer law, achieving over 70 \% average accuracy and 97\% semantic similarity. Unlike prior generic systems, our tool is uniquely structured for consumer cases (Section \ref{sec:methodology}).

\item \textbf{Consumer Decision Assist Tool (Similar Case Predictor)}, A sector-guided similar case judgment predictor integrates CoT-based sector classification with semantic and lexical similarity retrieval to achieve over 75\% accuracy, with its key innovation being the use of domain-specific sector filtering to enhance relevance in consumer law.
(Section \ref{sec:methodology}).

\item \textbf{CCFMS Dataset}, The first curated consumer law dataset in India with 152 case files and summaries authored by humans, specifically addressing consumer dispute resolution (Section \ref{sec:Dataset}).

\item \textbf{Prompt-based Automatic Evaluation Framework}, A novel 8-metric, part-wise evaluation framework using prompt-based automatic scoring, introducing domain-adapted metrics that strongly align with human judgments (Section \ref{sec:Evaluation}).
\end{enumerate}

\section{Related Work}
Legal summarization aims to condense complex legal texts, such as court rulings, legislative documents, and contracts, into accessible formats without losing critical legal meaning. Approaches typically include extractive, abstractive, and hybrid methods \cite{shukla-etal-2022-legal, zhang2024systematicsurveytextsummarization, akter2025comprehensivesurveylegalsummarization}. Recent advances focus on transformer-based models, which have significantly improved summarization accuracy in complex domains, yet domain-specific applications, particularly in Indian consumer law, remain scarce. Several Indian legal datasets like IL-TUR \cite{joshi-etal-2024-il} and MultiEURLEX \cite{chalkidis-etal-2021-multieurlex} support legal reasoning tasks but lack dedicated consumer law coverage. Case retrieval and similar case prediction have been widely studied in civil, criminal, and human rights domains \cite{wu-etal-2023-precedent, cui2023survey}, often leveraging legal element extraction to enhance relevance matching \cite{zongyue2023leeclegalelementextraction, deng-etal-2024-element}. Unsupervised retrieval methods using event extraction have also shown promise \cite{joshi-etal-2023-u}, but consumer case retrieval, especially in the Indian context, is underexplored.\\
Recent studies emphasize the importance of prompt engineering to maximize LLM performance in legal tasks \cite{sahoo2024systematic, wei2022chain}, with techniques like Chain of Thought (CoT) prompting facilitating multi-step reasoning. Evaluation of summarization quality has traditionally relied on human judgment , although concerns over reproducibility and scalability persist . Emerging work positions LLMs as reliable, reference-free evaluators \cite{liu-etal-2023-g, chiang-lee-2023-large, zheng2023judging, siledar-etal-2024-one}, offering scalable alternatives aligned with human assessments. Despite these advances, there remains a gap in building comprehensive summarization and retrieval systems tailored to Indian consumer law.

\begin{figure}
    \centering
    \includegraphics[width=0.9\linewidth]{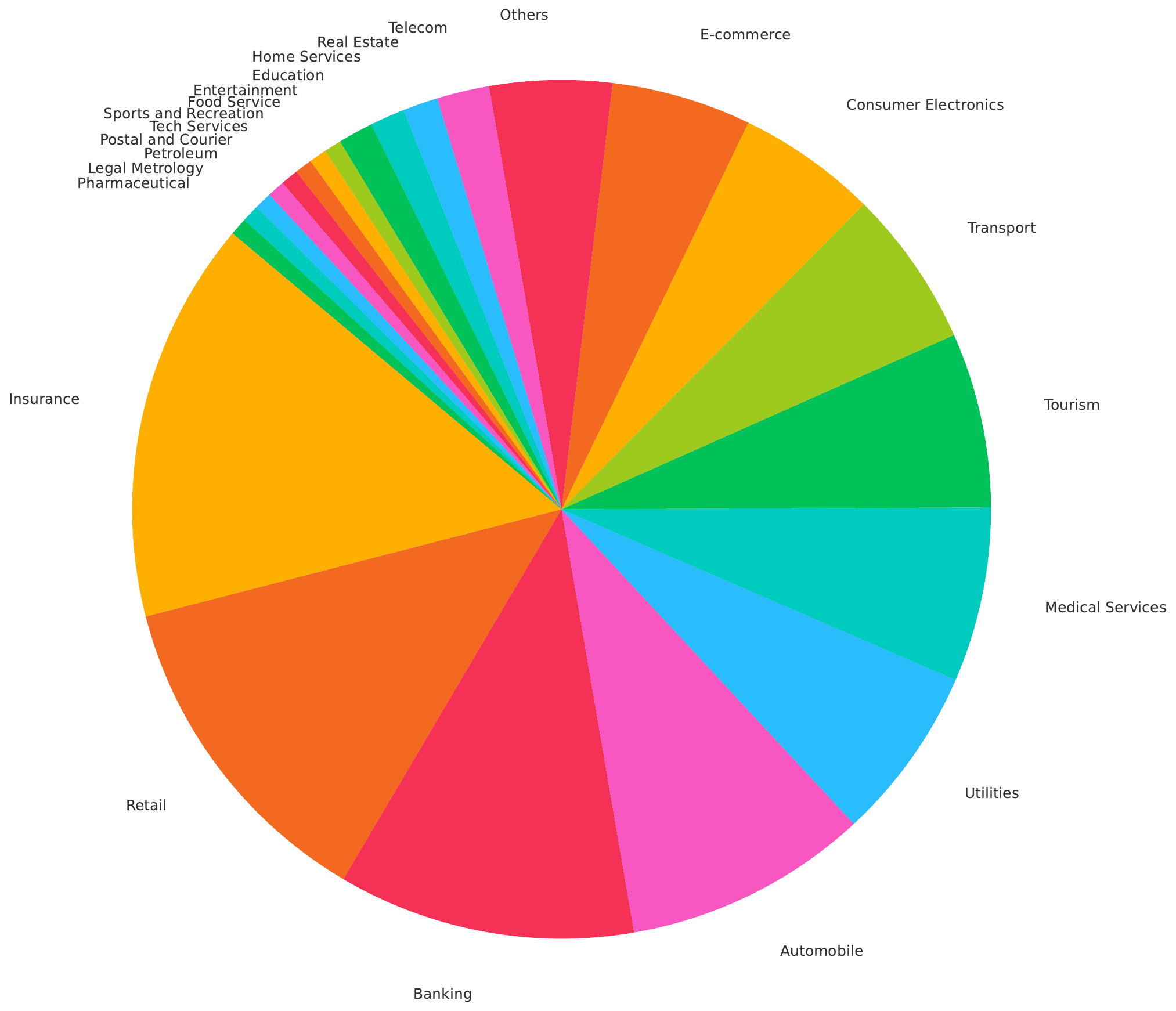}
    \caption{Distribution of consumer cases}
    \label{fig:enter-label}
\end{figure}

\section{Dataset}
\label{sec:Dataset}
For our present task, we propose the CCFMS dataset and use the Consumer case database \cite{ganatra2025textitgrahaknyayconsumergrievanceredressal}.
\subsection{CCFMS dataset}
The Consumer Case Files and Material Summaries (CCFMS) dataset comprises 152 carefully curated consumer case files across 23 diverse sectors, including banking, insurance and automobile, offering a comprehensive view of consumer-related disputes, claims, complaints, and legal issues.  Accompanying these case files are expert-created material summaries that concisely capture the six key elements (discussed in sec. \ref{sec:methodology}) essential for understanding each case, along with the five most similar case files. An example of a material summary is given in Appendix \ref{sumex}.

\begin{table*}[h!]
    \centering
    \scriptsize
    \captionsetup{justification=centering, width=\linewidth}
    \begin{tabular}{lccccc}
        \toprule
        \textbf{Models} & \textbf{Rouge-1} & \textbf{Rouge-2} & \textbf{Rouge-L} & \textbf{Bleu-1} & \textbf{BertScore} \\ 
        \midrule
        Llama 3.1 8B (Single prompt) &
49.07 & 23.41 & 26.36 & 28.13 & 96.16 \\ 
Llama 3.1 8B + Partwise + SR & \textbf{54.01} & 24.34 & 23.80 & \textbf{37.28} & 97.18
 \\ 
        Llama 3.1 8B + Partwise + CoT & 53.85 & \textbf{27.40} & \textbf{26.89} & 37.05 & \textbf{97.32} \\ 
        Ministral 8B + Partwise + CoT & 48.43 & 24.36 & 24.38 & 28.40 & 96.73 \\ 
        Deepseek 8B + Partwise + CoT & 45.01 & 16.30 & 18.45 & 29.89 & 96.14 \\
        Qwen 2.5 7B + Partwise + CoT  & 41.53 & 21.44 & 23.42 & 20.08 & 97.28 \\ 
        \bottomrule

    \end{tabular}

\caption{Performance comparison of generated summaries from different LLMs using ROUGE, BLEU, and BERTScore in reference-based evaluation. SR means Simple Restructured Prompt.}
\label{tab:bleu}
\end{table*}

\subsection{Consumer Case Judgement Database}
The consumer case database \cite{ganatra2025textitgrahaknyayconsumergrievanceredressal} contains 570 scraped case files from various sectors, including e-Commerce, telecommunications, healthcare, automobile, banking, real estate and travel. 
In our case, we have a gold-label sectoral mapping for all the case files, which enabled full accuracy. 
More details about the dataset can be found in \cite{ganatra2025textitgrahaknyayconsumergrievanceredressal}. 
These relevant cases are intended to help consumer forums make decisions. Extracts from annotated parts of a judgement with a brief are present in Appendix \ref{jury}.
\begin{table*}[h]
    \centering
    \scriptsize
    \resizebox{\textwidth}{!}{
    \begin{tabular}{@{}lcccccccc@{}}
        \toprule
        \textbf{Model Name} & \textbf{Over. Acc.} & \textbf{Oversim.} & \textbf{Over. Retr.} & \textbf{Iss. Acc.} & \textbf{Evid. Acc.} & \textbf{Iss. Form.} & \textbf{Sect. Rel.} & \textbf{Rel. Acc.} \\
        \midrule
        Llama 3.1 8B (Single Prompt) & 3.11 & 3.23 & 2.94 & 2.76 & 0.16 & 0.75 & 0.33 & 0.33 \\
        Llama 3.1 8B + Partwise + SR & \textbf{4.35} & 4.05 & \textbf{3.25} & 3.00 & 0.50 & \textbf{0.80} & 0.55 & 0.73 \\
        Llama 3.1 8B + Partwise + CoT & 4.25 & \textbf{4.19} & 3.14 & \textbf{3.50} & \textbf{0.67} & 0.67 & 0.60 & 0.75 \\
        DeepSeek 8B + Partwise + CoT & 3.30 & 3.35 & 2.71 & 3.43 & \textbf{0.67} & 0.71 & \textbf{0.95} & 0.48 \\
        Ministral 8B + Partwise + CoT & 4.15 & 3.95 & 3.05 & 3.25 & 0.50 & 0.70 & 0.70 & 0.10 \\
        Qwen 2.5 7B + Partwise + CoT & 4.25 & 4.10 & 3.00 & 3.20 & 0.40 & 0.75 & 0.70 & \textbf{0.85} \\
        \bottomrule
    \end{tabular}
    }
    \caption{Human evaluation of summaries generated by different models with Chain-of-Thought (CoT) prompting. The first four columns are rated on a 5-point Likert scale: Overview Accuracy, Oversimplification, Overview Retrieval, and Issues Accuracy. The last four columns are binary metrics: Evidence Accuracy, Issue Formatting, Sector Relevance, and Relief Accuracy. SR means Simple Restructured Prompt.}
    \label{tab:humaneval1}
\end{table*}

\begin{figure*}[htbp]
    \centering
    \includegraphics[width=\linewidth]{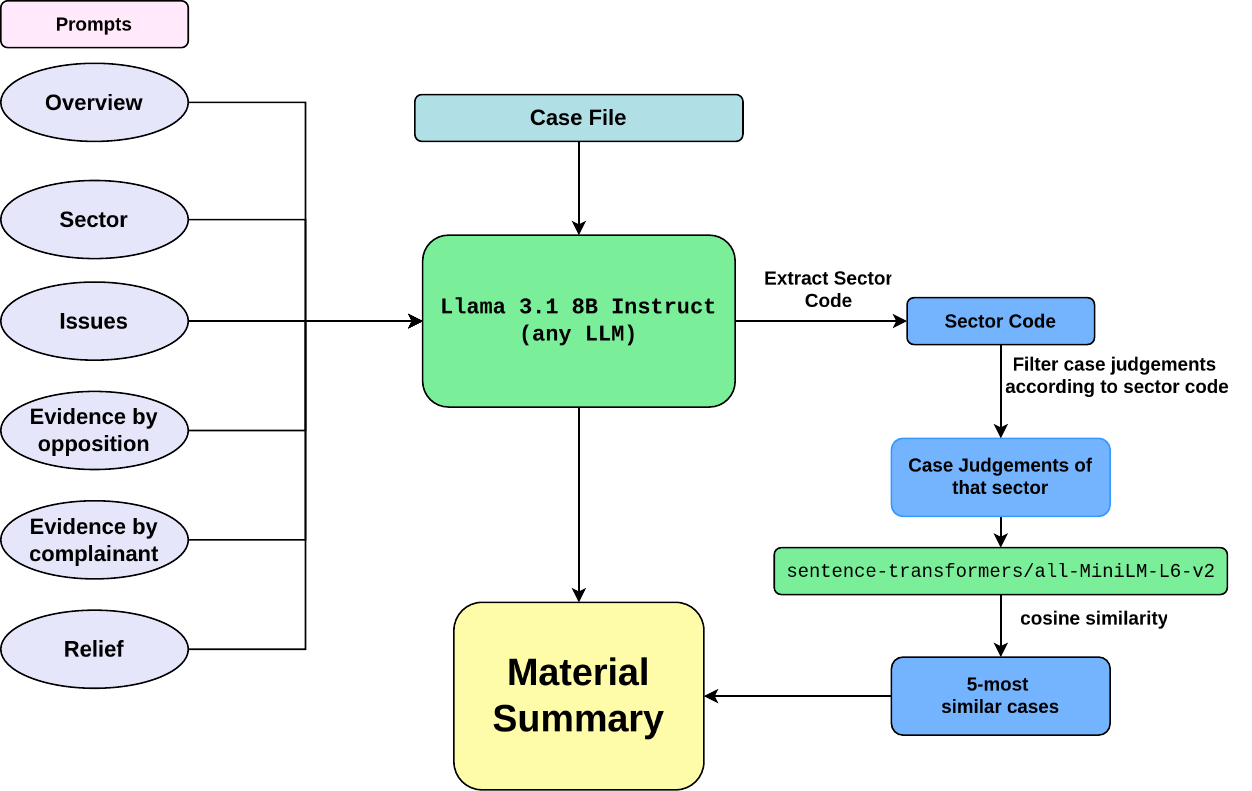}
    \caption{Detailed Architecture of NyayDarpan}
    \label{fig:architecture1}
\end{figure*}

\section{Methodology}
\label{sec:methodology}
Summarization is the act of distilling out a representative from the data \cite{zhang2024systematicsurveytextsummarization}.
Material summary generation involves distilling out information related to specific legal points.
The extraction of material summary has been performed using LLMs in a two-step process (Figure \ref{fig:architecture1}), which includes extraction and summarisation of six salient parts of the material summary from the case files and finding the 5 most similar consumer cases to this case file.\\
For the generation we used the following LLMs: \texttt{Llama 3.1 8B Instruct}\footnote{\url{https://huggingface.co/meta-llama/Llama-3.1-8B-Instruct}}, \texttt{DeepSeek R1 Distill Llama-8B}\footnote{\url{https://huggingface.co/deepseek-ai/DeepSeek-R1-Distill-Llama-8B}}, \texttt{Ministral 8B Instruct}\footnote{\url{https://huggingface.co/mistralai/Ministral-8B-Instruct-2410}} and \texttt{Qwen2.5 7B Instruct}\footnote{\url{https://huggingface.co/Qwen/Qwen2.5-7B-Instruct}}.
\subsection{Summary Structure}
The input to our system is a consumer case file comprising the complaint document and the written statement. These contain all relevant textual information, including the parties involved, specific claims, and evidence submitted by both sides. The system prompt, combined with the case file, is fed into the LLM, which generates a structured material summary with the following key components:
\begin{itemize}
\item \textbf{Overview:} A brief factual summary covering the disputed product or service, the grievance (defects, service deficiencies, failures), damages or inconvenience faced, any grievance mechanisms used (like prior complaints), the opposite party’s response, and the core legal issue.
\item \textbf{Sector:} Classification of the complaint into a predefined consumer protection sector using standard codes (e.g., Banking and Financial Services: 101) to ensure correct regulatory mapping.
\item \textbf{Issues:} A numbered list of key factual and legal issues raised by the complainant and counterarguments by the opposite party, focusing on defectiveness, consumer rights violations, and justification for compensation.
\item \textbf{Evidence:} Separate lists of evidence from both parties, where the complainant may provide receipts, contracts, or communication records, and the opposite party may submit warranties, service reports, or policy documents.
\item \textbf{Reliefs:} A summary of the remedies sought, including refunds, replacements, compensation, or reimbursement of legal costs.
\end{itemize}

\subsection{Extraction of Salient Parts}
The extraction of salient parts from consumer case files is implemented using a system prompt specifically designed for structured material summarization with LLMs. This process leverages the model's capability to identify and extract critical components from case documents through carefully engineered prompts. The architecture and step-wise methodology are detailed in the following subsections.

\subsubsection{System Prompt Construction}
A comprehensive system prompt is designed to extract six key components of a material summary: Overview, Sector, Issues, Evidence by Complainant, Evidence by Opposing Party, and Reliefs (see Figures \ref{fig:prompt-summary1} and \ref{fig:prompt-summary2}). The prompt includes precise definitions and clear instructions to ensure that the summaries adhere to a standardized structure.

To enhance performance, Nyay-Darpan employs a structured, part-wise prompting strategy instead of a single, monolithic prompt. The summary generation task is divided into six distinct sub-prompts, each focusing on one of the summary components. This targeted approach improves the semantic precision and coherence of the extracted parts and is optimized by varying the token limits based on the expected length of each section.

The effectiveness of this method is validated through comparative results (Table \ref{tab:bleu}), which show significant improvements in ROUGE, BLEU, and BERTScore metrics when compared to baseline methods.

To further enhance performance, we experimented with two prompting techniques: Simple Restructuring and Chain of Thought (CoT) Prompting.

The \textbf{Simple Restructuring} technique involves rewriting the original instructions to make them more explicit, direct, and logically segmented, with prompts divided into smaller, step-by-step tasks that guide the model in extracting specific parts of the summary. This clear, structured approach eliminates redundancy and ambiguity, significantly improving the accuracy and completeness of outputs, as illustrated in Figures \ref{fig:prompt-summary3}, \ref{fig:prompt-summary4}, and \ref{fig:prompt-summary5}. In contrast, the \textbf{Chain of Thought (CoT)} Prompting method guides the model through a step-by-step reasoning process, encouraging it to 'think aloud' by considering and validating each intermediate step before proceeding. This enhances the logical coherence and reduces errors in the generated summaries, with examples provided in Figures \ref{fig:prompt-COT1}, \ref{fig:prompt-COT2}, and \ref{fig:prompt-COT3}.

A comparative illustration of both prompting styles is provided in Figure \ref{fig:appendix-prompts-simple}.

\subsection{Similar Case Prediction}
In order to strengthen the decision-support functionality of Nyay-Darpan, a robust similar case prediction module is incorporated. The process begins with sector classification, performed using a Chain of Thought (CoT) prompt (Figure \ref{fig:prompt-COT1}). Once the sector is identified, similar case retrieval is executed by measuring semantic similarity between the current case's overview and historical case briefs within the same sector.

This is achieved by generating dense embeddings for the current case overview and past judgments using the transformer model: all-MiniLM-L6-v2\footnote{\url{https://huggingface.co/sentence-transformers/all-MiniLM-L6-v2/discussions}}. Cosine similarity is then computed to rank the most relevant historical judgments. In addition to semantic retrieval, BM25-based \cite{li2024bmxentropyweightedsimilaritysemanticenhanced} lexical retrieval is also employed to capture surface-level textual overlaps. Further, a hybrid retrieval approach that combines both BM25 and embedding-based similarity scores is implemented to ensure comprehensive retrieval that balances both lexical and semantic relevance.

This similarity-based retrieval framework ensures contextual relevance and provides legal practitioners with quick access to precedent cases, supporting more informed and consistent decision-making.

\begin{table}[htbp!]
    \centering
    \small
    \renewcommand{\arraystretch}{1.3} 
    \begin{tabular}{lccc}
        \hline
        \textbf{Model Name} & \textbf{Embed.} & \textbf{bm-25} & \textbf{Hybrid}\\ 
        \hline
        Llama 3.1 8B & 0.55 &0.56&0.44 \\ 
        Llama 3.1 8B + CoT & 0.60 &0.62 &0.52  \\ 
        DeepSeek 8B + CoT & \textbf{0.76} &\textbf{0.79} &\textbf{0.65}\\ 
        Ministral 8B + CoT & 0.59 &0.61 &0.50 \\ 
        Qwen 2.5 7B + CoT & 0.57 &0.59 &0.48\\ 
        \hline
    \end{tabular}
    \caption{Precision accuracy of different models with and without Chain-of-Thought (CoT) prompting.}
    \label{tab:model_performance}
\end{table}

\begin{table*}[h]
    \centering
    \scriptsize
    \resizebox{\textwidth}{!}{
    \begin{tabular}{@{}lcccccccc@{}}
        \toprule
        \textbf{Model Name} & \textbf{Over. Acc.} & \textbf{Oversimp.} & \textbf{Over. Retr.} & \textbf{Iss. Acc.} & \textbf{Evid. Acc.} & \textbf{Iss. Form.} & \textbf{Sec. Rel.} & \textbf{Rel. Acc.} \\
        \midrule
        Llama 3.1 8B (Single Prompt)  & 2.65 & 2.29 & 2.02 & 2.06 & 0.14 & 0.61 & 0.28 & 0.33 \\
        Llama 3.1 8B + Partwise + SR      & 3.53 & \textbf{4.17} & 2.90 & 3.53 & 0.33 & 0.50 & 0.63 & 0.60 \\
        Llama 3.1 8B + Partwise + CoT & \textbf{4.20} & 4.03 & \textbf{3.03} & \textbf{3.83} & \textbf{0.37} & \textbf{0.67} & 0.60 & \textbf{0.70} \\
        DeepSeek 8B + Partwise + CoT  & 3.23 & 3.03 & 2.37 & 3.67 & 0.33 & 0.57 & \textbf{0.90} & 0.27 \\
        Ministral 8B + Partwise + CoT & 3.57 & 3.87 & 2.70 & 3.67 & 0.37 & 0.50 & 0.67 & 0.10 \\
        Qwen 2.5 7B + Partwise + CoT  & 4.07 & 3.63 & 2.60 & 3.63 & 0.13 & 0.33 & 0.77 & 0.63 \\
        \bottomrule
    \end{tabular}
    }
    \caption{LLM-based evaluation of summaries generated by different models using \texttt{gpt-4o-mini}. The first four columns are rated on a 5-point Likert scale: Overview Accuracy, Oversimplification, Overview Retrieval, and Issues Accuracy. The last four columns are binary metrics: Evidence Accuracy, Issue Formatting, Sector Relevance, and Relief Accuracy. SR means Simple Restructured Prompt.}
    \label{tab:gpt1}
\end{table*}

\section{Evaluation and Results}
\label{sec:Evaluation}
We evaluate the generated summaries using reference-based, reference-free, and human evaluation methods. Eight metrics, recommended by legal experts, assess the quality and correctness. Metrics are evaluated using either a 5-point Likert scale or a binary scale.

The evaluation metrics are:
\begin{enumerate}
\item \textbf{Overview Accuracy}: Assesses how faithfully the summary captures key factual details like dates, amounts, parties, and major facts. Higher scores indicate greater accuracy.
\item \textbf{Overview Oversimplification}: Evaluates whether essential elements such as product/service, issues, damages, grievance mechanisms, and claims are retained. Lower scores indicate omissions or excessive simplification.
\item \textbf{Overview Retrieval}: Measures the extent to which the summary retrieves critical facts from the original case. Higher scores reflect comprehensive coverage.
\item \textbf{Sector Relevance}: Checks if the sector name and code correctly match the human-annotated material summary. Binary evaluation (Yes/No).

\item \textbf{Issues (Formatting)}: Verifies that issues are presented in a structured, numbered format, clearly distinguishing claims from both parties. Binary evaluation (Yes/No).

\item \textbf{Issues (Accuracy)}: Measures whether the identified issues are factually correct and logically derived from the case. Higher scores indicate greater accuracy.

\item \textbf{Evidence Accuracy}: Ensures the evidence aligns with the original case, without hallucinations or omissions. Binary evaluation (Yes/No).

\item \textbf{Relief Accuracy}: Verifies that the reliefs stated match those in the original case. Binary evaluation (Yes/No).

\end{enumerate} 
\subsection{Reference-based lexical and semantic evaluation}
We evaluated the performance of our summarization model using ROUGE, BLEU, and BERTScore (\citet{zhang2020bertscoreevaluatingtextgeneration}). ROUGE scores (ROUGE-1, ROUGE-2, and ROUGE-L) were used to measure n-gram overlap between generated and reference summaries. (see Table \ref{tab:bleu} for scores and Appendix \ref{sec:appendix}.1 for details of packages used)
\begin{table}[h]
    \centering
    \begin{tabular}{lc}
        \toprule
        \textbf{Metric} & \textbf{Spearman Correlation} \\
        \midrule
        Overview Accuracy        & 0.5105  \\
        Oversimplification         & 0.5181  \\
        Overview Retrieval         & 0.4804  \\
        Issues Accuracy            & 0.4282  \\
        Evidence Accuracy          & 0.7134 \\
        Issue Formatting           & 0.7886  \\
        Sector Relevance           & 0.8551  \\
        Relief Accuracy            & 0.6986  \\
        \bottomrule
    \end{tabular}
    \caption{Spearman's rank correlation coefficient of human evaluation with LLM-based evaluation using \texttt{gpt-4o-mini} model as an evaluator of the generated summaries.}
    \label{tab:spearman_corr}
\end{table}

\subsection{Evaluation of summaries on 8-point metrics}

In human evaluation, we achieve an average score of more than 4 (out of 5) on the overview accuracy, oversimplification, overview retrieval and issue accuracy metrics, and a score of more than 0.60 out of 1 on the Evidence Accuracy, Issue Formatting, Sector Relevance, and Relief Accuracy metrics, demonstrating the general effectiveness of using CoT with the \texttt{Llama-3.1-8B-Instruct} model (Table \ref{tab:humaneval1}). We use \texttt{gpt-4o-mini} model for the LLM-based evaluation (Table \ref{tab:gpt1}). The correlation result of LLM-based evaluation with human evaluation is in Table \ref{tab:spearman_corr}.  Appendix \ref{prompts} presents the prompts used for LLM-based evaluation. The same prompts are also meant as instructions for annotators to facilitate the evaluation process (Appendix \ref{anno}).

\subsection{Evaluation of Similar Case Prediction}
Out of the 5 judgments predicted, a team of legal experts checks each of the judgments to ensure which are relevant to a case file and which are not. Table \ref{tab:model_performance} gives accuracy in terms of precision.
\section{Observations and Analysis}
\begin{itemize}
    \item \textbf{Sector classification was key for similar judgement prediction}, with Deepseek-8B achieving the best results and directly driving similar case prediction performance (see Tables \ref{tab:humaneval1}, \ref{tab:model_performance}, \ref{tab:gpt1}).

    \item \textbf{Banking and insurance were the most confusing sectors}, where all models, including Deepseek-8B, often misclassified cases (see Appendix Figure \ref{fig:confusing-example}), suggesting a need for better disambiguation.

    \item \textbf{Llama 3.1 8B outperformed Deepseek-8B on summarization tasks} like overview and relief accuracy (see Tables \ref{tab:humaneval1}, \ref{tab:gpt1}), showing its strength in fact-based extraction despite weaker reasoning.

    \item \textbf{The hybrid retrieval method was the best overall}, combining lexical and semantic similarity to outperform single-feature approaches in similar case prediction.
\end{itemize}

\section{Conclusion and Future Work}
In this paper, we presented our decision assist tool as a summarizer cum similar case predictor for assisting the Indian consumer law forums, quasi-judicial bodies, as well as for customers by summarizing case files and gathering similar case files for speedy and perfect resolution of disputes. We evaluated our approach against other state-of-the-art models, but found our simple prompting approach to be at par or better. In terms of automatic evaluation, our model performs decently. For future work, we plan to use more prompting techniques to improve the performance of our algorithms. We also plan to use other techniques for clustering so as to improve the similar case prediction of the tool.

\section*{Limitations}
The process of generating a material summary, while effective in structuring case details, faces several limitations. First, the approach relies on the quality and completeness of the input case documents. Missing or ambiguous information leads to incomplete summaries. Similar case prediction depends on available case data, and limited and lower-quality case files hinder accuracy. Additionally, variations in judicial reasoning and jurisdiction-specific nuances can impact the relevance of predicted cases. Finally, while structured summaries improve readability, they may oversimplify complex legal arguments, potentially omitting critical contextual details. Future improvements could integrate more robust techniques with higher-quality data to enhance accuracy and adaptability.

\section*{Ethical Considerations}
The CCFMS dataset was created by a team of legal experts who carefully curated case files to ensure accurate representation of consumer disputes. The dataset development followed ethical guidelines to maintain fairness, confidentiality, and neutrality in summarizing legal proceedings.  

To generate material summaries, we employed a system prompt designed to extract structured information from case files. While this approach enhances consistency and objectivity, potential ethical risks exist. These include the risk of misinterpretation due to inherent biases in legal language processing in the domain of consumer law and the possibility of oversimplifying complex legal arguments. Additionally, system-generated summaries must be evaluated critically to ensure they do not inadvertently favour any party in a dispute.  

We encourage the research and legal communities to use this framework responsibly. Further refinements, including expert-in-the-loop evaluations and expansion of the dataset with more quality examples, can help mitigate biases and improve reliability in legal case summarization in the domain of consumer law.

\bibliography{custom}
\bibliographystyle{acl_natbib}
\appendix

\section{Appendix}
\label{sec:appendix}

\subsection{Inference Hyperparameters and Evaluation Libraries}

We used the following hyperparameters during inference. The \texttt{max\_new\_tokens} parameter was varied depending on the expected length of the output: sector name and number were typically concise (approximately 16 tokens), reliefs asked required more descriptive detail (around 256 tokens), and case overviews and issues demanded longer outputs (up to 512 tokens) to retain key facts and context. Other decoding hyperparameters were set as follows: \texttt{temperature} = 0.7, \texttt{top\_p} = 0.95, and \texttt{top\_k} = 50. All inference was conducted using the vLLM engine\footnote{\url{https://docs.vllm.ai}}. 

We used the following evaluation libraries and models to assess the quality of generated outputs. ROUGE scores were computed using Google's \texttt{rouge\_score} library\footnote{\url{https://github.com/google-research/google-research/tree/master/rouge}}. BLEU scores were calculated with the \texttt{sentence\_bleu} function from the \texttt{nltk.translate.bleu\_score} module\footnote{\url{https://www.nltk.org/_modules/nltk/translate/bleu_score.html}}. For semantic similarity evaluation, we used BERTScore via the \texttt{bert-score} library with the \texttt{bert-base-uncased} model\footnote{\url{https://pypi.org/project/bert-score}}. These metrics collectively provided surface-level and semantic-level assessments of the generated text.
For similar case prediction, the hybrid method gives 50\% weightage to lexical (bm25) features and 50\% to semantic (embedding) features.

\subsection{Summary Example}
\label{sumex}
{\ttfamily
\begin{flushleft}
MATERIAL SUMMARY EXAMPLE\\

\textbf{Overview:}\\
The complainant purchased an iPhone from an authorised seller of Apple, which turned out to be defective from the very first day. Even after visiting the authorised service centre of Apple, the phone was not repaired. A replacement of the phone was provided, which also started to face software and hardware issues, and the same could not be fixed by the service centre, so the phone was subsequently returned to the customer. The Opposite Party contended that, as no exact defect could be identified by the authorised service centre, the product could not fall under warranty. However, the OP replaced the product. But even after satisfactory replacement, frivolous complaints were made as contended by Apple. Aggrieved by the response from Apple, the complainant has filed the complaint seeking to get the price of the defective phone along with compensation.\\

\textbf{Sector \& Code}: Consumer Electronics, 110

\textbf{Issues:}\\
- Whether the complainant is a ‘consumer’ of Apple?\\
- Whether the sale of a defective product along with failure to repair such defect amounts to deficiency in service?\\
- Whether the defective product was well within the terms and conditions of warranty?\\
- Whether the complaint was frivolous and the opposite party is entitled to any relief against it?

\textbf{Evidence -- Complainant:}\\
CE1: ID proof\\
CE2: Purchase bill\\
CE3: Delivery report\\
CE4: Letter from the opponent\\
CE5: Bill of the new phone

\textbf{Evidence -- Opposite Party:}\\
OPE1: Copy of Apple’s one-year limited warranty\\
OPE2: Evidence by way of affidavit on behalf of OP no. 1 filed on 10th March 2019, written argument on 10/11/2020

\textbf{Reliefs Sought:}\\
- Refund of Rs. 18,740/- with interest at the rate of 18\% per annum from the day of loss till the realization of payment or replace it with a new piece of iPhone.\\
- Compensation of Rs. 30,000/- to the complainant for the mental harassment and Rs. 20,000/- as cost of the present legal proceeding.
\end{flushleft}
}

\subsection{Annotated Judgement}
\label{jury}
 Judgement Name:- Leno Lhouvisier Zinyü vs. The Chairman, Max Life Insurance Company Ltd. and Ors.,\\
 Citation:- CC/1/2015 2023 SCDRC Nagaland\\
 Sector Name:- Insurance \\
 Sector Code:- 102\\
 Brief:- In this case, where neither the insurer nor the insured come to the commission with clean hands (which is also the case in the present one), the commission held that it will be in the interest of justice to restore the parties back to the position they were before the contract.

 \subsection{Annotator background and instructions}
 \label{anno}
 Human evaluation was conducted by legal experts on 150 summaries generated by each model. Evaluators were provided with detailed written guidelines outlining the evaluation criteria and the structure of the summaries. Given the time-intensive nature of expert evaluation, we conducted a single round of annotation. As a result, inter-annotator agreement metrics were not computed. We acknowledge this as a limitation and plan to include multi-rater evaluations with agreement analysis in future work.

 The human experts involved in both dataset annotation and evaluation were legal practitioners, academicians, and law graduates/postgraduates with relevant expertise in consumer protection law. The process was overseen by a senior professor from a top Indian law school. To maintain quality and avoid bias, the annotation, evaluation, and instruction design teams were kept independent. Detailed written guidelines were provided at every stage, including definitions for the six summary components and eight evaluation metrics, with clear instructions on structure, prioritization, and judgment criteria.

\subsection{Prompts}
\label{prompts}
The detailed system prompts for generation, as well as the evaluation prompts, are attached in the following pages of the appendices of this paper.

\onecolumn

\begin{figure}[htbp!]
    \centering
    \begin{minipage}{0.45\textwidth}
        \textbf{Example of Simple Restructuring Prompt}
        
        \begin{tcolorbox}[colback=gray!5!white,colframe=gray!75!black,title=Prompt: Simple Restructuring]
        \textbf{Objective:} Extract key information from a product review.  
        
        \textbf{Task Segmentation:}
        \begin{enumerate}
            \item \textbf{Product Name:} Identify the name of the product being reviewed.  
            \item \textbf{Reviewer Opinion:} Summarize the main opinion of the reviewer (positive/negative/neutral).  
            \item \textbf{Pros Highlighted:} List the positive points mentioned.  
            \item \textbf{Cons Mentioned:} List the negative points mentioned.  
            \item \textbf{Overall Rating:} Extract the overall rating if available.  
        \end{enumerate}
        
        \textbf{Expected Output:}  
        The output should be formatted as follows:  
        
        \texttt{
        \textbf{Product Name:} [Name of the product] \\
        \textbf{Reviewer Opinion:} [Positive/Negative/Neutral] 
        \textbf{Pros:} [List of positive aspects] \\
        \textbf{Cons:} [List of negative aspects] \\
        \textbf{Overall Rating:} [Rating value]
        }
        \end{tcolorbox}
    \end{minipage}%
    \hspace{0.05\textwidth} 
    \begin{minipage}{0.45\textwidth}
        \textbf{Example of CoT-based Prompting}
        
        \begin{tcolorbox}[colback=gray!5!white,colframe=gray!75!black,title=Prompt: CoT-based Prompting]
        \textbf{Objective:} Analyze a product review to determine if the product is recommended.  
        
        \textbf{Step-by-Step Reasoning:}
        \begin{enumerate}
            \item \textbf{Step 1:} Identify the main sentiment of the reviewer (positive or negative).  
            \item \textbf{Step 2:} List the main reasons supporting the sentiment.  
            \item \textbf{Step 3:} Check if the reasons are mostly related to product quality, service, or pricing.  
            \item \textbf{Step 4:} Determine if the reviewer explicitly recommends or discourages the product.  
            \item \textbf{Step 5:} Conclude if the product is generally recommended based on sentiment and reasons.  
        \end{enumerate}
        
        \textbf{Expected Output:}  
        The output should be formatted as follows:  
        
        \texttt{
        \textbf{Step 1:} [Main sentiment identified] \\
        \textbf{Step 2:} [List of reasons] \\
        \textbf{Step 3:} [Category of reasons: Quality/Service/Pricing] \\
        \textbf{Step 4:} [Recommendation status: Yes/No] \\
        \textbf{Step 5:} [Final conclusion: Recommended/Not Recommended]
        }
        \end{tcolorbox}
    \end{minipage}
    \caption{Comparison of Simple Restructuring and CoT-based Prompting Techniques on Product Review Analysis}
    \label{fig:appendix-prompts-simple}
\end{figure}

\begin{figure*}
\centering
\tcbset{}
\begin{tcolorbox}[width=\textwidth]
\begin{Verbatim}
Overview:

The complainant purchased an iPhone from an authorised seller of Apple which
turned out to be defective from the very first day. Even after visiting the 
authorized service center of Apple, the phone was not repaired. A replacement of 
the phone was provided which also started to face software and hardware issues 
and the same could not be fixed by the service center and the phone was 
subsequently returned to the customer.

The Opposite Party contended that as no exact defect could be identified by the 
authorized service center, the product could not fall under warranty. However, 
the OP replaced the product. But even after satisfactory replacement, frivolous 
complaints were made as contended by Apple.

Aggrieved by the response from Apple, the complainant has filed the complaint
seeking to get the price of the defective phone along with compensation.

SECTOR AND SECTOR CODE:
Consumer Electronics, 110

ISSUES:
1. Whether the complainant is a ‘consumer’ of Apple?
2. Whether the sale of a defective product along with failure to repair such
defect amounts to deficiency in service?
3. Whether the defective product was well within the terms and conditions of 
warranty?
4. Whether the complaint was frivolous and the opposite party is entitled to 
any relief against it?

EVIDENCE PRESENTED BY THE COMPLAINANT:
CE1: ID proof of applicant
CE2: Bill of disputed mobile
CE3: Delivery report
CE4: Copy of letter issued by the opponent
CE5: Copy of bill of new mobile

EVIDENCE PRESENTED BY THE OPPOSITE PARTY:
OPE1: Copy of Apple’s one-year limited warranty
OPE2: Evidence by way of affidavit on behalf of OP no. 1 filed on 10th March 
2019 written argument on 10/11/2020

RELIEF:
1. Refund of Rs. 18,740/- with interest at the rate of 18% per annum from the
day of loss till the realization of payment or replace it with a new piece.
2. Compensation of Rs. 30,000/- to the complainant for the mental harassment 
and Rs. 20,000/- as cost of the present legal proceeding.
\end{Verbatim}
\end{tcolorbox}
\caption{Consumer Case Example: iPhone Defect Dispute}
\label{fig:consumer-case-example}
\end{figure*}

\begin{figure*}
\centering
\tcbset{}
\begin{tcolorbox}[width=\textwidth]
\begin{Verbatim}
Task Description:
Evaluate the accuracy of the issues presented in the generated summary by
comparing it is with the ground truth of the legal case summary. Ensure that the 
issues align with the scope and factual details provided in the ground truth. 
The issues must be logically derived from the factual matrix and the claims made 
in the case. Inaccuracies, omissions, or misalignments should result in a lower 
score based on the evaluation criteria.

Ground truth summary:
{original}

Generated Summary:
{generated}

Evaluation Criteria:
Rate the accuracy of the issues on a scale from 1 to 5:

<score>5</score>: The issues are perfectly accurate, comprehensive, and 
logically derived from the facts and claims.
<score>4</score>: The issues are mostly accurate, with minor inconsistencies 
or omissions.
<score>3</score>: The issues are somewhat accurate but include some significant
inconsistencies or omissions.
<score>2</score>: The issues are largely inaccurate or fail to align with the
factual details.
<score>1</score>: The issues are completely inaccurate, irrelevant, or not 
derived from the factual matrix.
Instructions:


Instructions:
1. Assign a score strictly based on the evaluation criteria.  
2. Include the score within `<score></score>` tags at the end of your response.  

Response Format:
Final score: Present the score in this format: `<score>[1-5]</score>`.  

\end{Verbatim}
\end{tcolorbox}
\caption{Prompt for LLM-based evaluation of Issues Accuracy metric}
\end{figure*}

\begin{figure}[ht]
\centering
\tcbset{}
\begin{tcolorbox}
\begin{Verbatim}
Task Description:
You are tasked with evaluating the accuracy of the generated summary by 
comparing it with the ground truth of legal case summary. Your primary goal is 
to assess how well the generated summary reflects the factual content, including
critical details such as dates, amounts, events, facts, and parties involved. 
Accuracy is paramount, and any incorrect or misleading information should lead 
to a lower score based on the provided criteria.

Ground truth summary:
{original}

Generated Summary:
{generated}


Evaluation Criteria:
Score from 1 to 5 based on accuracy:
<score>5</score>: Perfectly accurate; no factual inaccuracies or misleading
details.  
<score>4</score>: Mostly accurate; contains minor factual errors or slightly 
misleading details.  
<score>3</score>: Moderately accurate; some factual inaccuracies, but key 
information remains intact.  
<score>2</score>: Significantly inaccurate; contains major errors or misleading 
details but retains some correct facts.  
<score>1</score>: Highly inaccurate; major errors or misleading details severely 
distort the facts of the case.  

Instructions:
1. Assign a score strictly based on the evaluation criteria.  
2. Include the score within `<score></score>` tags at the end of your response.  

Response Format:
Final score: Present the score in this format: `<score>[1-5]</score>`.  

\end{Verbatim}
\end{tcolorbox}
\caption{Prompt for LLM-based evaluation of Overview Accuracy metric}
\end{figure}

\begin{figure}[ht]
\centering
\tcbset{}
\begin{tcolorbox}
\begin{Verbatim}
Task Description:
You are tasked with evaluating the level of oversimplification of the generated
summary by comparing it with the ground truth of legal case summary. 
Specifically, assess whether the generated summary includes and adequately
describes the following critical components:
The service or product in question.
The problem with the product or service.
The damage caused by the problem.
The grievance mechanisms that have been used.
The claims made by the opposite party.
The parties involved in the issue.

If the summary omits any of these components or oversimplifies them, assign a
lower score based on the criteria below.

Ground truth summary:
{original}

Generated Summary:
{generated}


Evaluation Criteria:
Score the level of oversimplification from 1 to 5:
<score>5</score>: All key elements are present and clearly described without
oversimplification.
<score>4</score>: Most key elements are included, with minor omissions or slight
oversimplifications.
<score>3</score>: Some key elements are omitted or overly simplified, but the 
main aspects are still represented.
<score>2</score>: Many important elements are omitted or significantly 
oversimplified, leading to a vague summary.
<score>1</score>: Critical elements are missing or severely oversimplified, 
distorting the essence of the case.


Instructions:
1. Assign a score strictly based on the evaluation criteria.  
2. Include the score within `<score></score>` tags at the end of your response.  

Response Format:
Final score: Present the score in this format: `<score>[1-5]</score>`.  
\end{Verbatim}
\end{tcolorbox}
\caption{Prompt for LLM-based evaluation of Overview Oversimplication metric}
\end{figure}

\begin{figure}[ht]
\centering
\tcbset{}
\begin{tcolorbox}
\begin{Verbatim}
Task Description:
You are tasked with evaluating how well the genrated summary retrieves relevant
facts incomparison with ground truth summary. Assess whether the generated 
summary includes all critical facts and details present in the ground truth. 
Any missing or inaccurately represented facts should result in a lower score 
based on the criteria provided.

Ground truth summary:
{original}

Generated Summary:
{generated}

Evaluation Criteria:
Rate the summary's ability to retrieve relevant facts on a scale from 1 to 5:

<score>5</score>: The summary retrieves all critical facts with no omissions 
or inaccuracies.
<score>4</score>: The summary is accurate but misses a few minor details.
<score>3</score>: Several important facts are missing, though the summary 
retains some critical details.
<score>2</score>: Many significant facts are missing or inaccurately represented,
reducing clarity.
<score>1</score>: The summary fails to retrieve critical facts or entire 
sections of the original case file.

Instructions:
1. Assign a score strictly based on the evaluation criteria.  
2. Include the score within `<score></score>` tags at the end of your response.  

Response Format:
Final score: Present the score in this format: `<score>[1-5]</score>`.  
\end{Verbatim}
\end{tcolorbox}
\caption{Prompt for LLM-based evaluation of Overview Retrieval metric}
\end{figure}

\begin{figure}[ht]
\centering
\tcbset{}
\begin{tcolorbox}
\begin{Verbatim}
Task Description:
Review the sector relevance in the generated summary by comparing it with the 
ground truth of legal case summary. Compare the sector name in the generated 
summary with the sector name in the ground truth. If sector name matches and 
is relevant, mark the evaluation as "Yes." If either is incorrect or missing, 
mark it as "No."

Ground truth summary:
{original}

Generated Summary:
{generated}

Instructions:
Assign 'Yes' or 'No' strictly based on the evaluation criteria. Provide a 
detailed explanation justifying the score. Include the final score using
<score></score> tags.

Response Format Example:
Provide a detailed explanation of the evaluation.
Final score: Score - <score>Yes</score> or <score>No</score>.
\end{Verbatim}
\end{tcolorbox}
\caption{Prompt for LLM-based evaluation of Sector Relevance metric}
\end{figure}

\begin{figure}[ht]
\centering
\tcbset{}
\begin{tcolorbox}
\begin{Verbatim}

Task Description:
Review the evidence section in the generated summary by comparing it with the
ground truth of legal case summary. Verify whether the list of evidence matches
the evidence provided in the ground truth summary. Ensure there is no 
hallucinated evidence and that all mentioned evidence corresponds accurately to
the ground truth.

Ground truth summary:
{original}

Generated Summary:
{generated}

Evaluation Criteria:
Yes: The evidence in the generated summary matches the ground truth summary,
with no hallucinated or missing evidence.
No: There are discrepancies, such as hallucinated evidence or missing 
references from the ground truth summary.

Instructions:
Assign 'Yes' or 'No' strictly based on the evaluation criteria. Provide a 
detailed explanation justifying the score. Include the final score using
<score></score> tags.

Response Format Example:
Provide a detailed explanation of the evaluation.
Final score: Score - <score>Yes</score> or <score>No</score>.
\end{Verbatim}
\end{tcolorbox}
\caption{Prompt for LLM-based evaluation of Evidence Accuracy metric}
\end{figure}

\begin{figure}[ht]
\centering
\tcbset{}
\begin{tcolorbox}
\begin{Verbatim}

Task Description:
Evaluate whether the issues in the generated summary are captured in the correct 
format. Specifically, check if:
1. The issues are presented as a numbered list.
2. Each issue addresses a distinct question of fact.
3. The factual claims by the complainant and those contested by the opposing 
party are clearly stated.

Ground truth summary:
{original}

Generated Summary:
{generated}

Evaluation Criteria: Does the formatting meet the criteria?): [Yes/No]

Instructions:
Assign 'Yes' or 'No' strictly based on the evaluation criteria. Provide a 
detailed explanation justifying the score. Include the final score using
<score></score> tags.

Response Format Example:
Provide a detailed explanation of the evaluation.
Final score: Score - <score>Yes</score> or <score>No</score>.


\end{Verbatim}
\end{tcolorbox}
\caption{Prompt for LLM-based evaluation of Issue Formatting metric}
\end{figure}

\begin{figure}[ht]
\centering
\tcbset{}
\begin{tcolorbox}
\begin{Verbatim}

Task Description:
Review the relief section in the generated summary. Check if the relief 
presented in the generated summary match those mentioned in the ground truth 
summary.

Ground truth summary:
{original}

Generated Summary:
{generated}

Evaluation Criteria:
Yes: The relief section in the generated summary matches the ground truth 
summary, with no hallucinates or missing relieves.
No: There are discrepancies, such as hallucinated relieves or missing relieves 
from the ground truth summary.

Instructions:
Assign 'Yes' or 'No' strictly based on the evaluation criteria. Provide a 
detailed explanation justifying the score. Include the final score using
<score></score> tags.

Response Format Example:
Provide a detailed explanation of the evaluation.
Final score: Score - <score>Yes</score> or <score>No</score>.


\end{Verbatim}
\end{tcolorbox}
\caption{Prompt for LLM-based evaluation of Relief Accuracy metric}
\end{figure}

\begin{figure}[ht]
    \centering
    \tcbset{}
    \begin{tcolorbox}
        \begin{Verbatim}
Extract the following 6 components of the material summary and no other headings.
Every material summary should contain only these 6 components and 
no other headings.

1. Overview: In this section, include a description of the facts of the 
given consumer case.
The factual summary you prepare should include the following: 
what was the service or product in question which forms the subject of the 
consumer grievance?; 
what was the problem with the product or service?; 
What damage was caused by the problem?;
what is/are the grievance mechanism(s) that have been availed by the consumer 
thus far, if any, and 
what is the claim of the opposite party? Clearly specify the parties 
in the dispute, especially if there are multiple parties. 
A longer list of opposite parties (over 4) may be condensed into a short 
summary of opposite parties. You can end this by mentioning the core of the 
legal issue being disputed in one sentence. This section should be at least
7-10 lines long.
 
2. Sector: What sector of consumer grievance/protection does this case fall 
under from the list below? The list of sectors is as follows: Extract the sector 
along with the number next to it. The sectors can only be one of the following
with their respective sector codes:-
Banking and Financial Services  101 Insurance   102
Retail - Clothing   103 Retail - Electronics    104
Retail - Home & Furniture   105
Retail - Groceries and FMCG 106 Retail - Beauty & Personal Care 107
E-commerce  108 Telecommunications  109 
Medical Services (including Negligence) 112
Transport - Airlines    113 Transport - Railways    114 Real Estate 115
Utilities (Electricity, Water)  116 Automobiles 117 Food Services   118 
Education   120
Entertainment and Media 121 Legal Services  122 Home Services   123 
Sports and Recreation   124 Technology Services 125
Legal Metrology 126 Petroleum   127 Postal and Courier  128 Others  999

3. Issues: This section of the material summary should primarily be the issues
brought before the court. 
Include a numbered list of the issues in the case, i.e., what factual claims 
have been put forth by the complainant and which are contested by the 
opposing party. 
Each issue should represent a distinct question. 



        \end{Verbatim}
    \end{tcolorbox}
    \caption{Prompt for material summary generation - part 1}
    \label{fig:prompt-summary1}
\end{figure}

\begin{figure}[ht]
    \centering
    \tcbset{}
    \begin{tcolorbox}
        \begin{Verbatim}
4. Evidence presented by the complainant: A list of the evidentiary material 
[e.g., purchase receipts, contracts, tickets, bills, photos, videos], 
if mentioned in the copy of the complaint that has been filed before the court 
by the complainant, with a brief description of each item. The list should be 
numbered preceded in the following style: 
"CE1. [mention a brief description of the first item of complainant evidence]
CE2. [mention a brief description of 
the second item of complainant party evidence]
CE3. [mention a brief description of the third item of complainant 
evidence, and so on]."

If the complaint doesn't explicitly mention evidence, consider phrases 
like "evidence attached as annexure" to indicate supporting documentation. 
If the evidence list is not provided in the complaint 
copy, write "Nil" in the Material Summary in this section.

5. Evidence presented by the opposite party: A list of the evidentiary material
[e.g., purchase receipts, contracts, tickets, bills, photos, videos], 
if mentioned in the copy of the written statement,
that has been filed before the court by the opposite party, with a brief 
description of each item. The list should be numbered preceded 
in the following style: 
"OPE1. [mention a brief description of the first item of opposite party 
evidence]
OPE2. [mention a brief description of the second item of opposite party 
evidence]
OPE3. [mention a brief description of the third item of opposite 
party evidence, and so on]."

If the complaint doesn't explicitly mention evidence, consider phrases 
like "evidence attached as annexure" to indicate supporting documentation.
If the evidence list is not provided in the written statement copy, 
write "Nil" in the Material Summary in this section.

6. Reliefs: In this section, include a numbered list of reliefs requested by 
the complainant in the prayer of the complaint copy. It should be a numbered 
list of reliefs claimed, with the figures if mentioned.

        \end{Verbatim}
    \end{tcolorbox}
    \caption{Prompt for material summary generation - part 2}
    \label{fig:prompt-summary2}
\end{figure}

\begin{figure}[ht]
    \centering
    \tcbset{}
    \begin{tcolorbox}
        \begin{Verbatim}
Prompts for extraction of each individual part is given below

1. Overview:- Extract a detailed overview of the consumer case from the 
provided case file.Your output should follow this format:
Overview: [Write the overview here in a single paragraph]
The overview must include the following information:
What is the product or service that is the subject of the consumer grievance?
What specific issue or defect did the consumer face with the product or service?
What was the impact or damage caused to the consumer?
What steps or grievance mechanisms (if any) has the consumer already used?
What is the claim or response made by the opposite party or parties?
Clearly identify the parties involved in the dispute. If there are more than 
four opposite parties, provide a short summary or grouping instead of listing 
all names.
Conclude with a single sentence summarizing the core legal issue in dispute.
The answer should be in a single paragraph and should be at least 7–10 lines 
long to ensure completeness and clarity.
 
2. Sector: From the given case file, identify the sector name and sector code 
that best represents the subject of the consumer grievance. Your classification 
should be based on two main factors: The product or service involved in the 
dispute The identity or nature of the opposite party (e.g., a bank, hospital,
airline,e-commerce platform, etc.) Use this combined information to determine
the most appropriate sector. Your output should strictly follow this format: 
Sector:- [Sector Name], [Sector Code] 
Do not include any explanation or reasoning. 
Select only one sector name and code from the list below: 
Banking and Financial Services 101 Insurance 102
Retail - Clothing 103 Retail - Electronics 104 Retail - Home & Furniture 
105 Retail - Groceries and FMCG 106 Retail - Beauty & Personal Care 
107 E-commerce 108 Telecommunications 109 Consumer Electronics 
110 Healthcare and Pharmaceuticals 111 Medical Services (including Negligence) 
112 Transport - Airlines 113 Transport - Railways 114 Real Estate 
115 Utilities (Electricity, Water) 116 Automobiles 
117 Food Services 118 Travel and Tourism 119 Education 120 
Entertainment and Media 
121 Legal Services 122 Home Services 123 Sports and Recreation 
124 Technology Services 125 Legal Metrology 126 Petroleum 
127 Postal and Courier 128 Others 999


        \end{Verbatim}
    \end{tcolorbox}
    \caption{Prompts for simple restructuring - part 1}
    \label{fig:prompt-summary3}
\end{figure}
\begin{figure}[ht]
    \centering
    \tcbset{}
    \begin{tcolorbox}
        \begin{Verbatim}
Prompts for extraction of each individual part is given below


3. Issues: Extract the key issues presented in the case file.
These should reflect the disputed questions or factual claims that have been 
brought before the court.
Each issue must be a specific point of contention between the complainant and 
the opposing party—claims made by 
the complainant and denied or challenged by the opposite party.
The output should follow this format:
Issues:-
[First issue] [Second issue] ...
Ensure each issue is clearly worded and focused on one distinct question 
or claim. Only include issues that are actively disputed or form part of the 
legal conflict. Do not include any explanatory or background information.

        \end{Verbatim}
    \end{tcolorbox}
    \caption{Prompts for simple restructuring - part 2}
    \label{fig:prompt-summary4}
\end{figure}
\begin{figure}[ht]
    \centering
    \tcbset{}
    \begin{tcolorbox}
        \begin{Verbatim}
Prompts for extraction of each individual part is given below
4. Evidence by the complainant:-
Extract the evidence presented by the complainant from the case file.
These are the items of evidentiary material (such as receipts, contracts, 
tickets, bills, photos, videos, etc.) that are mentioned in the complaint copy 
filed before the court.
Present the output strictly in the following format:
Evidence presented by the complainant:-
CE1. [Brief description of the first evidence item]
CE2. [Brief description of the second evidence item]
CE3. [Brief description of the third evidence item]
(...continue as needed)
Use the prefix “CE” followed by the number for each item.
Only include evidence explicitly mentioned in the complaint copy.
Do not include anything outside this format—no explanations, headers, or 
summaries
 
5.Evidences by the opposite party:-
Extract the evidences presented by the opposite party from the case file.
These are the items of evidentiary material (such as receipts, contracts, 
tickets, bills, photos, videos, etc.) that are mentioned in the written 
statement filed by the opposite party before the court.
Present the output strictly in the following format:
Evidences presented by the opposite party:-
OPE1. [Brief description of the first evidence item]
OPE2. [Brief description of the second evidence item]
OPE3. [Brief description of the third evidence item]
(...continue as needed)
Use the prefix “OPE” followed by the number for each item.Only include 
evidence explicitly mentioned in the case file or written statement.
Do not include anything outside this format. No commentary, no headers, no 
summaries—just the list as shown.

6. Extract the reliefs requested by the complainant from the case file.
These are the reliefs mentioned in the prayer section of the complaint copy.
Present the output in the following format:
Reliefs:-
[First relief requested, include figures if mentioned]
[Second relief requested]
[Third relief requested]
(...and so on)
Do not include anything else—only the numbered list as shown.
No explanations or extra text.
        \end{Verbatim}
    \end{tcolorbox}
    \caption{Prompts for simple restructuring - part 3}
    \label{fig:prompt-summary5}
\end{figure}

\begin{figure}[ht]
    \centering
    \tcbset{}
    \begin{tcolorbox}
        \begin{Verbatim}
Prompts for extraction of each individual part is given below
1. Carefully read the provided consumer case file and think step-by-step to 
extract a comprehensive overview. Start by identifying:
The product or service that is central to the grievance.
Next, describe the specific defect or issue the consumer experienced with it.
Then, consider what impact, harm, or inconvenience it caused the consumer.
Examine whether the consumer has tried any grievance mechanisms or escalation 
steps (e.g., complaints, repairs, refund requests).
Analyze the response or counterclaims made by the opposite party or parties.
Clearly identify the parties involved in the case. If there are more than four 
opposite parties, group or summarize them to maintain clarity.
Finally, reflect on the above details and summarize the core legal issue in 
dispute in one sentence.
Now, write a single detailed overview paragraph (7–10 lines minimum) 
incorporating all of the above points.
Format your response as:
Overview: [Write the full overview paragraph here]
 
2. Sector: Carefully examine the provided case file and think step-by-step
to identify the correct sector classification.
First, determine the product or service that is central to the grievance.
Then, analyze the identity or nature of the opposite party — what type of 
organization or business are they? (e.g., a bank, hospital, e-commerce site).
Use both the product/service and the opposite party’s nature to assess which 
sector best fits.
Refer to the list of sectors and select the single most appropriate match based 
on the combined information.
Do not explain your choice — only output the final classification in the 
required format.
Your response must strictly follow this format (no extra text or explanation):
Sector:- [Sector Name], [Sector Code]
Use only one of the following predefined sectors:
Banking and Financial Services 101  
Insurance  102 Retail - Clothing  103  
Retail - Electronics  104  Retail - Home & Furniture  105  
Retail - Groceries and FMCG 106  Retail - Beauty & Personal Care 107
E-commerce  108  Telecommunications  109 Consumer Electronics 110 
Healthcare and Pharmaceuticals 111  
Medical Services (including Negligence) 112 Transport - Airlines    113  
Transport - Railways  114  Real Estate 115  
Utilities (Electricity, Water)  116  Automobiles 117  Food Services   118  
Travel and Tourism 119 Education   120  Entertainment and Media 121  
Legal Services  122  Home Services   123  Sports and Recreation   124  
Technology Services 125  Legal Metrology 126  
Petroleum   127  Postal and Courier  128  Others  999 


        \end{Verbatim}
    \end{tcolorbox}
    \caption{COT Prompts for summarization - part 1}
    \label{fig:prompt-COT1}
\end{figure}

\begin{figure}[ht]
    \centering
    \tcbset{}
    \begin{tcolorbox}
        \begin{Verbatim}
Prompts for extraction of each individual part is given below

3. Issues:- Carefully read the case file and follow these reasoning steps to
extract the key legal issues in dispute:
Identify the claims made by the complainant—what specific allegations, 
factual assertions, or complaints have they raised?
Next, analyze the responses or counterclaims made by the opposite party—what 
parts of the complainant’s case do they deny, reject, or challenge?
For each area of disagreement, formulate a precise, specific issue that reflects
a point of contention between the parties.
Make sure each issue captures only one distinct claim or factual dispute.
Exclude any background details, narrative summaries, or uncontested facts.
Present your final answer in this strict format:
Issues:-
1) [First issue]
2) [Second issue]
...(and so on)
Only include actively disputed issues that form part of the legal conflict.

 
4. Evidence by the complainant: Carefully examine the complaint copy filed
by the complainant and follow these steps to extract the evidentiary items:
Scan through the text to identify any explicit references to physical or 
digital materials submitted as part of the complaint.
Look for items such as receipts, invoices, tickets, contracts, bills, emails, 
letters, photographs, videos, or any other documents cited by the complainant.
Ensure that each item is mentioned in the complaint itself and is part of the 
official submission before the court.
For each evidence item, write a brief but clear description, focusing only on 
its type and relevance.
Do not include items implied but not mentioned, or any interpretation, 
background, or legal commentary.
Output your answer strictly in this format:
Evidence presented by the complainant:-
CE1. [Brief description of the first evidence item]
CE2. [Brief description of the second evidence item]
CE3. [Brief description of the third evidence item]
(...continue as needed)
Do not include anything outside this format.
 

        \end{Verbatim}
    \end{tcolorbox}
    \caption{COT Prompts for summarization - part 2}
    \label{fig:prompt-COT2}
\end{figure}

\begin{figure}[ht]
    \centering
    \tcbset{}
    \begin{tcolorbox}
        \begin{Verbatim}
Prompts for extraction of each individual part is given below

5. Evidence by the opposite party: Carefully read the written statement or reply
filed by the opposite party in the case file and follow these 
steps to extract the evidence they have presented:
Identify all explicitly mentioned documents or materials submitted by the 
opposite party as part of their defense or response.
Look for references to bills, receipts, contracts, photographs, videos, 
official records, letters, emails, or any other material intended to support 
their version of events. Verify that each item is specifically mentioned in the
written statement or attached as supporting material by the opposite party.
For each valid evidence item, write a concise and factual description, 
limited to what is stated in the file.
Do not include any inferred evidence, commentary, or background explanation.
Output your answer strictly in the following format:
Evidence presented by the opposite party:-
OPE1. [Brief description of the first evidence item]
OPE2. [Brief description of the second evidence item]
OPE3. [Brief description of the third evidence item]
(...continue as needed)
Do not include anything beyond the list. No summaries, no headings, no 
reasoning—just the formatted output.

6. Relief:- Follow these reasoning steps to extract the reliefs requested:
1. Locate the prayer or relief section of the complaint, usually found at the 
end of the complaint copy.  
2. Identify each specific request made by the complainant to the court — this 
could include refunds, compensation, damages, interest, litigation costs,
or any declaratory or injunctive relief.  
3. Ensure that each relief is explicitly mentioned in the prayer and not 
inferred from the narrative.  
4. If a monetary amount is stated, include the figure as written.  
5. List each relief as a separate bullet point, without interpretation, summary, 
or rephrasing.
Present your final answer strictly in this format:  
Reliefs:-
[First relief requested, include figures if mentioned]  
[Second relief requested]  [Third relief requested](...and so on)
Do not include any additional explanation, headings, or commentary—only the
relief list.
        \end{Verbatim}
    \end{tcolorbox}
    \caption{COT Prompts for summarization - part 3}
    \label{fig:prompt-COT3}
\end{figure}

\begin{figure}[ht]
    \centering
    \tcbset{}
    \begin{tcolorbox}
        \begin{Verbatim}
The complainant, Gurraya S/o Basayya, is a retired employee of Opposite Party
No. 2, Assistant Provident Fund Commissioner. Opposite Party No. 1 (OP 1) and
Opposite Party No. 3 (OP 3) are the Provident Fund authorities. The Complainant 
was a member of the Family Pension Scheme 1971 during employment, which was 
replaced by the Employees' Pension Scheme 1995 of which he became a continued 
member. Upon his retirement in 2000, his monthly pension was settled at Rs. 350
by OP 1.In 2016, the complainant came to know that his pension amount was 
calculated erroneously and was lesser than his entitlement. 
His representation to OP 1 for revision was denied. 

The OPs, especially OP 1, have denied any deficiency in calculating the 
complainant's pension as per the applicable provisions of the Employees' 
Pension Scheme 1995. OP 1 claims the complainant is not eligible for additional 
2-year weightage benefit as he did not complete the required 20 years of service 
or has not attained 58 years under the 1995 scheme itself. They also contend 
that the complainant is not a consumer and the complaint is grossly time-barred.
The complainant has thus filed this complaint alleging deficiency in service 
by the OPs.write latex for adding this as overview of a confusing example 
for sector prediction

        \end{Verbatim}
    \end{tcolorbox}
    \caption{Overview of the most confusing example for the sector}
    \label{fig:confusing-example}
\end{figure}

\end{document}